\documentclass[lettersize,journal]{IEEEtran}
\IEEEoverridecommandlockouts
\usepackage{cite}
\usepackage{amsmath,amssymb,amsfonts}
\usepackage{dsfont} 
\usepackage{algorithmic}
\usepackage{graphicx}
\usepackage{epstopdf} 
\usepackage{textcomp}
\usepackage{xcolor}
\usepackage{dblfloatfix} 
\usepackage[utf8]{inputenc}
\usepackage[english]{babel}
\usepackage{xspace} 
\usepackage{algorithm}
\usepackage{subcaption}
\usepackage{amsthm} 
\usepackage{mathtools} 
\usepackage[normalem]{ulem}
\usepackage{mathtools}

\usepackage{array}
\newcolumntype{P}[1]{>{\centering\arraybackslash}p{#1}}
\newcolumntype{M}[1]{>{\centering\arraybackslash}m{#1}}

\usepackage{balance}
\usepackage{multicol}
\usepackage{enumerate}
\usepackage{enumitem}
\usepackage{verbatim}

\usepackage{tikz}
\usetikzlibrary{matrix,calc,shapes}
\tikzset{
  treenode/.style = {shape=rectangle, rounded corners,
                     draw, anchor=center,
                     text width=7em, align=center,
                     top color=white, font=\footnotesize, bottom color=blue!20,
                     inner sep=1ex},
  treenode2/.style = {shape=rectangle, rounded corners,
                     draw, anchor=center,
                     text width=10em, align=center,
                     top color=white, font=\footnotesize, bottom color=blue!20,
                     inner sep=1ex},
  decision/.style = {treenode2, diamond, aspect=3, text width=12em, inner sep=-3pt,  minimum height=2.1cm,minimum width=2.1cm},
  decision2/.style = {treenode2, diamond, aspect=3, inner sep=-4pt,  minimum height=2.1cm,minimum width=2.8cm},
  root/.style     = {shape=rectangle, rounded corners,
                     draw, anchor=center,
                     text width=4em, align=center,
                     top color=white, font=\footnotesize,, bottom color=red!30, text width=4.35em,inner sep=1ex},
  env/.style      = {treenode},
  finish/.style   = {root, bottom color=green!40, text width=4.35em},
  dummy/.style    = {circle,draw}
}
\newcommand{\yes}{edge node [above] {\small yes}}

\newcommand{\yesR}{edge node [right] {\small yes}}

\newcommand{\no}{edge  node [above]  {\small no}}
\newcommand{\noL}{--  node [left]  {\small no}}

    \newtheorem{note}{Note}

\newcommand{\suchthat}{\, \mid \,}
\newcommand{\suchthatbig}{\, \Huge{\mid} \,}

\def\BibTeX{{\rm B\kern-.05em{\sc i\kern-.025em b}\kern-.08em
    T\kern-.1667em\lower.7ex\hbox{E}\kern-.125emX}}

\newcommand{\blue}[1]{{\color{black}{#1}}}

\newcommand{\aBS}{$a\!B\!S$\xspace}
\newcommand{\aBSs}{$a\!B\!S$s\xspace}

\newcommand{\gNB}{$g\!N\!B$\xspace}
\newcommand{\gNBs}{$g\!N\!B$s\xspace}

\catcode`\%=12
\newcommand\pcnt{
\catcode`\%=14

\setlength{\abovedisplayskip}{-2.25pt}
\setlength{\belowdisplayskip}{6pt}

\renewcommand{\baselinestretch}{1}

\begin{document}

\title{Exact Resource Allocation for Fair Wireless Relay}


\author{
        Edgar~{Arribas},~
        Vincenzo~{Mancuso},~
        Vicent~{Cholvi}
\IEEEcompsocitemizethanks{
\IEEEcompsocthanksitem E.~{Arribas} is with Universidad Cardenal Herrera-CEU, CEU Universities, Val\`encia, Spain, edgar.arribasgimeno@uchceu.es. V.~{Mancuso} is with IMDEA Networks Institute, Madrid, Spain,vincenzo.mancuso@imdea.org. V.~{Cholvi} is with Universitat Jaume I (UJI), Castell\'o, Spain, vcholvi@uji.es.
\IEEEcompsocthanksitem Work supported by AEON-CPS (TSI-063000-2021-38), funded by the Ministry of Economic Affairs and Digital Transformation of Spain and the European Union NextGeneration-EU in the framework of the Spanish Recovery, Transformation and Resilience Plan;
and by the grant INDI23/17 and GIR23/01 from Universidad Cardenal Herrera-CEU, CEU Universities.
}
}

\maketitle
\begin{abstract}
In relay-enabled cellular networks, the intertwined nature of network agents calls for complex schemes to allocate wireless resources.
Resources need to be distributed among mobile users while considering how relay resources are allocated, and constrained by the traffic rate achievable by base stations and over backhaul links.
In this letter, we derive \blue{an \textsl{exact}} resource allocation scheme that achieves $\pmb{\max}$--$\pmb{\min}$ fairness across mobile users, found with a linear complexity with respect to the number of mobile users \textsl{and} relays. \blue{The results reveal that the proposed scheme remarkably outperforms current solutions.}
\end{abstract}


\section{\blue{Introduction}}

\begin{figure}[t]
    \centering
    \vspace{-1.5mm}
    \includegraphics[trim={5mm 3mm 4mm 3mm}, width=0.752\linewidth]{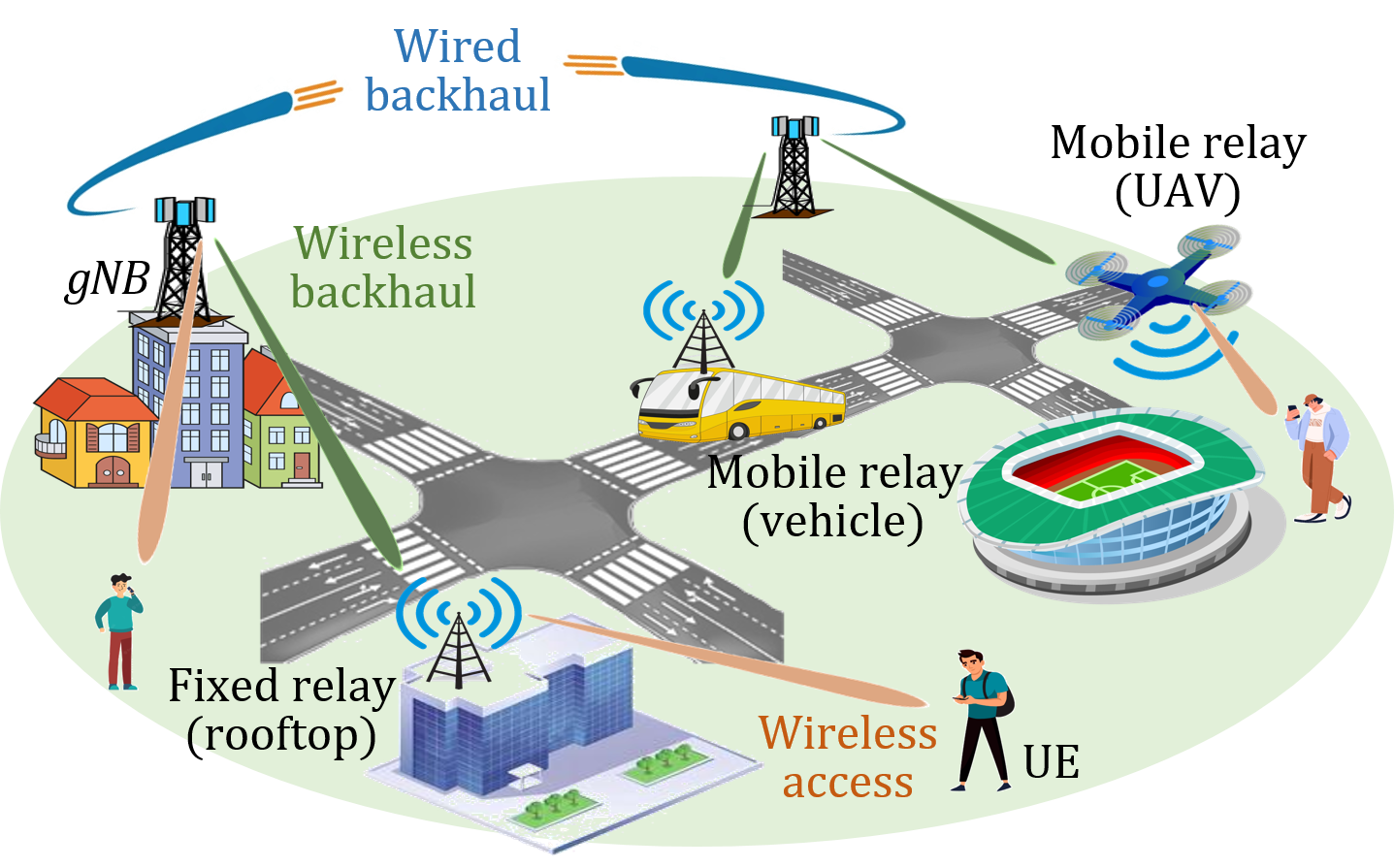}
    \vspace{-1mm}
    \caption{Reference scenario.}
    \vspace{-5mm}
    \label{fig:RefScenario}
\end{figure}

We consider a heterogeneous relay-enabled network~\cite{555c482195ae4754b9a93767b041053f} formed by a set of fixed \gNBs (\emph{Next Generation Node B}
)
providing wireless service both to mobile users and relays. Figure~\ref{fig:RefScenario} illustrates the considered scenario. It can be seen that there are two \gNBs that provide service to one mobile user and three relays (a rooftop tower, a bus and an unmanned aerial vehicle -- UAV). In turn, relays provide service to other mobile users (e.g., on the bus or in the stadium).

We derive a mechanism that provides a fair
rate allocation
to mobile users in downlink.
\blue{Specifically,}
we guarantee $\max$--$\min$ fairness~\cite{555c482195ae4754b9a93767b041053f},
\blue{i.e.,}
we maximize the performance of the worst-case
user, so potential service outages
\blue{are}
minimized.
\blue{Although}
alternative metrics exist for fairness,
in
this work, we
\blue{adopt}
$\max$--$\min$ because several practical systems require a minimum level of performance guarantees, below which the service cannot be properly deployed, hence customers would not pay for it. A wide range of data services fall into this category: online streaming and multimedia real-time applications, augmented reality, etc. The quality of these services does not improve linearly or with a continuous function of, e.g., bandwidth and delay, but rather experiences a staircase quality function with very few steps, which saturates at some level~\cite{10.1145/3339825.3391855}. For such services, what matters the most is to guarantee that all customers reach a level at which the service can be used.

The complexity of relay architectures makes the analysis quite difficult due to the intertwined nature of all the involved agents. Indeed, \gNB resources must be allocated not only to directly served users, but also shared with relays, and relays may reuse
wireless resources to serve their mobile users, thus generating interference.
Additionally, the use of \gNB resources is also constrained by the backhaul capacity.
Finally, 
wireless resources must be assigned quickly to be able to adapt to changing scenarios, as guaranteed by our proposal.

\vspace{-4mm}
\subsection*{\blue{Related Work}}
In the last years there has been an increasing number of studies focused on resource allocation in heterogeneous networks~\cite{555c482195ae4754b9a93767b041053f}.
\blue{Although $\max$--$\min$ resource allocation for \textsl{single} cells was optimally resolved in~\cite{coluccia2012optimality}, the extension of that problem to relay-aided networks is not trivial, and has been studied in different ways. Thus,}
here
we review
\blue{the available}
previous works, showing the different directions followed by them, \blue{and highlight how our work differs from existing results}.

In~\cite{8611177}, the authors focus on a downlink wireless network aided by a single UAV, which
\blue{aims}
to maximize the minimum average rate among all users.
In~\cite{9257576}, the authors investigate the use of the non-orthogonal multiple access technique for the case of a single UAV relay and solve a joint channel-and-power allocation problem with an iterative algorithm under $\max$--$\min$ fairness, yet they do not achieve optimal results.
\blue{Unlike our work, \cite{8611177} and~\cite{9257576}}
\blue{do not consider}
the case of multiple relays. 

In~\cite{moayedian2020fair}, the authors study proportional and $\max$--$\min$ fairness mechanisms in cognitive radio networks, where secondary users act as relays, aiming to provide acceptable rates. However, different from us, their analysis is restricted to IoT scenarios and
\blue{needs to solve}
non-convex problems, which prevents finding optimal results in reasonable time scales\blue{, while our approach finds exact solutions in linear time}.
In~\cite{DBLP:journals/telsys/ElgendyIE18}, the authors consider a scenario similar to ours. However, they take restrictive assumptions regarding how resources are allocated, and ignore inter-cell interference as well as interference between \gNBs and relays.
With that, they propose a \blue{suboptimal} heuristic and show that it can improve fairness.

In~\cite{tariq2022fairness}, the authors consider satellite-terrestrial relay networks in which rates are maximized under fairness constraints for user association and spectrum allocation.
However,
the complexity leads them to resort to heuristics that are suboptimal, unless infinite iterations are run, which results impractical. In~\cite{moghaddas2022relay}, authors address relay selection in dense heterogeneous networks to manage load balancing fairness, yet their focus is mainly oriented to device-to-device communications. 
\blue{However, with the approaches of~\cite{tariq2022fairness} and~\cite{moghaddas2022relay},}
a minimum service level for users cannot be guaranteed\blue{, different from what addressed in our work}.

Available \blue{works differ from our proposal in the sense that they either use just one relay, address different communication scenarios, or approach fairness in ways that cannot guarantee a minimum service performance, all of them ignoring in fact the presence of backhaul bottlenecks.} 

\vspace{-4mm}
\subsection*{\blue{Contributions}}
\blue{Novelty and contributions of this letter are as follows: 
\begin{itemize}
  \item We develop a $\max$--$\min$ fair resource allocation scheme for wireless relay networks that allows to jointly allocate resources to both mobile users \textsl{and} several relays, considering wired and wireless backhaul bottleneck constraints, which precludes the direct use of existing schedulers.
  \item Our algorithm finds the \textsl{exact} solution for the associated optimization, which goes beyond existing results.
  \item Such exact solution is found with linear complexity on the number of mobile users \textsl{and} relays, which is a strong advantage when it comes to practical implementations.
  \item The performance evaluation shows that our proposal remarkably outperforms current schemes when adapted to the framework of wireless relay networks, revealing that, actually, the scheme derived in this letter is needed.
\end{itemize}
}
\vspace{-4mm}

\section{System Model}

\blue{Table~\ref{tab:symbols_model} summarizes the system model parameters used in this letter.}
We consider a wireless relay-enabled network composed by a set
of fixed \gNBs and a set of relays that provide cellular service to a set of mobile users.
\blue{We model downlink traffic, i.e, traffic eventually delivered to mobile users by a \gNB or a relay.}
Each \gNB is attached to a wired backhaul network, whereas each relay is attached to one \gNB by means of a wireless backhaul link.
This represents a realistic framework for heterogeneous cellular networks that offers
 $(i)$ a flexible way to adapt to 
 occasional events and emergencies (e.g., from the case of crowded events to the case in which cellular coverage has to be temporary brought where no coverage is typically needed or because of an emergency or a specific ``mission'' requiring network support upgrades) and $(ii)$ an affordable way to extend network services without incurring the costs of a fixed infrastructure extension (e.g., when a ``volatile'' infrastructure is needed and the cost of a fixed one would not be otherwise recovered through the revenue associated with the service)~\cite{555c482195ae4754b9a93767b041053f}.

The set of relays attached to \gNB~$g$ is denoted as $\mathcal{R}_g$, the set of \gNB-served users is denoted as $\mathcal{U}_g$, for each \gNB $g$, the set of users served by relay $r$ is denoted as $\mathcal{U}_r$, and the set of users served by some relay attached to \gNB $g$ is denoted~as~$\mathcal{U}_g^*$.

Each  \gNB $g$ receives a maximum traffic capacity rate (denoted $\tau_g$) from the wired backhaul network, perhaps different from that of the other \gNBs. We denote as $W^g_\text{relays}$ the bandwidth of \gNB $g$ dedicated to relays and as  $W^g_\text{users}$ the bandwidth of \gNB $g$ dedicated to users directly attached to $g$.
In addition, each relay $r$ will allocate its bandwidth, which we denote as $W^r_\text{users}$, among the users it serves (note that $W^g_\text{relays}$, $W^g_\text{users}$ and $W^r_\text{users}$ are fixed values, since the assignment of spectrum bands to operators is performed by means of government auctions where only channels of fixed bandwidth are offered~\cite{Auction}). Such bands for mobile users and relays may be deployed by the operator as either orthogonal or reused bands. What matters for our analysis is that interference, if present, is accounted for.
After that, operators can split the assigned bandwidth into smaller portions to allocate sub-channels to specific groups of users and services, according to their target (e.g., optimize a fair network performance).

On another hand, it must be taken into account that practical systems cannot assign arbitrarily small bandwidth to individual stations or users~\cite{liu2019introduction}. Concretely, each relay obtains \blue{\sl at least}  $W^\text{min}_\text{relays}$, while each served mobile user receives
\blue{\sl at least} $W^\text{min}_\text{users}$.

\blue{Mobile users access downlink wireless resources with an OFDMA scheme, as for 3GPP mobile broadband networks~\cite{Auction}.
We assume that all \gNBs and relays use their entire available bandwidth, which
in practice, is the case that requires optimization. Hence, we consider that mean SINR (\emph{signal to interference \& noise ratio}) values are constant with respect to user resource allocation, and are solely determined by the inter-cell interference level, which in turn depends on which frequencies are used by \gNBs and relays. Instead, scheduling at the \gNB or relay prevents intra-cell interference.}


Although the above-mentioned interference can be reduced by making
\gNBs
use 3D-beamforming or adopting orthogonal frequencies, depending on the scenario it will be necessary to take into account the signal strength of each wireless channel, measured as the SINR.
We denote by $\gamma_{g,r}$ the SINR of the relay link between \gNB $g$ and a relay $r$, and by $\gamma_{s,u}$ the SINR of the access link between a station $s$ (either a \gNB or a relay) and a mobile user $u$. \blue{As wireless networks perform resource allocation based on the channel state information perceived
(basically, the SINR observed),
at the moment of distributing resources the scheduler is already aware of the users and relays cell selection and thus the SINR channel values, so that those $\gamma$ parameters need to be considered here as problem inputs.}


\begin{table}[t]
\vspace{-0.mm}
\centering
     \renewcommand{\arraystretch}{1.05} 
     \setlength{\tabcolsep}{3pt} 
\caption{System model parameters} 
\label{tab:symbols_model}
\vspace{-2mm}
{\begin{tabular}{|p{1.75cm}|p{6.5cm}|}
\hline
\textsl{Parameter} & \textsl{Description}   \\
\hline
$\mathcal{R}_g$  &
Set of relays attached to \gNB $g$. \\
\hline
$\mathcal{U}_g$, $\mathcal{U}_r$, $\mathcal{U}^*_g$   &
Mobile users attached to \gNB $g$, mobile users attached to a relay $r$, and mobile users attached to any relay $r$ (i.e., $\mathcal{U}^*_g=\bigcup_{r\in\mathcal{R}_g} \!{\cal U}_{r}$).  \\
\hline
$\tau_g$ & Maximum traffic rate of \gNB $g$. \\
\hline
$W^g_\text{relays}$, $W^g_\text{users}$, $W^r_\text{users}$ & Bandwidth of \gNB $g$ dedicated to relays,  bandwidth of \gNB $g$ dedicated to mobile users  and bandwidth of relay $r$ (dedicated to mobile users). \\
\hline
$W^\text{min}_\text{relays}$, $W^\text{min}_\text{users}$ & Minimum bandwidth for each relay and mobile user.\\
\hline
$\gamma_{s,y}$ & SINR between $s$ and $y$, where $s$ is a station (a \gNB or a relay) and $y$ is either a mobile user or a \blue{relay}.\\
\hline
\end{tabular}}
\vspace{-5mm}
\end{table}

\vspace{-2mm}
\section{The Resource Allocation Problem}

The aim of our work is to optimize the $\max$-$\min$ fairness of the throughput received by mobile users. This is not a trivial task, as all the involved agents (\gNBs, relays and mobile users) are intertwined (e.g., resources of mobile users from one relay cannot be allocated without knowing what backhaul resources that relay will get, depending on other relay resources and the \gNB bottleneck over the wired backhaul), while the interference management also involves different types of colliding wireless channels.
Since at resource allocation the network disposes of the \emph{channel state information} (CSI) feedback necessary to know the SINRs of the channels, each \gNB will be able to solve the resource allocation problem for its relays, its mobile users, and the users of relays attached to that \gNB in a concurrent and independent manner, by~using~the~convex program that we will introduce next in~\eqref{eq:CP}.

More formally, it will be necessary to obtain, for each relay~$r$ and for each mobile user $u$, both the share of bandwidth assigned (denoted $w_r$ and $w_u$), and the throughput experienced by the network node (denoted $T_r$ and $T_u$).

%
%

In~\eqref{eq:CP} we formulate, for each \gNB $g$, the corresponding resource allocation optimization in a Convex Program (CP):

\small
\begin{equation}
\hspace{-1.mm}
  \begin{cases}
    \hspace{3.0cm}\mathclap{\max \min \left \{T_u \suchthatbig u \in \mathcal{U}_g \bigcup \; \mathcal{U}^*_g  \right\}, \quad  \text{s.t.: }}  \\ \vspace{-1.5mm} \\

    1.~  w_r \geq W^{\min}_\text{relays},  & \forall r \!\in\! \mathcal{R}_g;  \\

   2.~ \sum_{r\in\mathcal{R}_g} \! w_r = W^g_\text{relays}; & \\

   3.~ T_r \!\leq\! w_r  \log_2\!\left(1 \!+\! \gamma_{g, r}\right), & \forall r \!\in\! \mathcal{R}_g;  \\ \vspace{-1.5mm} \\

    4.~ w_u \geq W^{\min}_\text{users}, & \forall  u\!\in \mathcal{U}_g\bigcup\mathcal{U}_g^{*}; \\

    5.~ \sum_{u\in\mathcal{U}_g} \! w_u = W^g_\text{users}; & \\


    6.~ \sum_{u \in \mathcal{U}_r} \! w_u = W^r_{\text{users}}, & \forall  r \!\in\! \mathcal{R}_g; \\ \vspace{-1.5mm} \\

    7.~ T_u \!\leq\!  w_u \!\log_2\!\left(1 \!+\! \gamma_{s,u}\right)\!, & \forall (s, u) \!\in\! \left(\!\{g\} \bigcup \mathcal{R}_g\!\right) \!\times\!  \left(\mathcal{U}_g \bigcup \mathcal{U}^*_g\!\right)\!; \!\!\!\!\!\!\!\!\!\!\!\!\!\! \\

    8.~ \sum_{u \in \mathcal{U}_r} \! T_u \!\leq  T_r, & \forall  r \!\in\! \mathcal{R}_g; \\

    9.~ \!\sum_{u\in\mathcal{U}_g}\limits \! T_u \!+ \! \sum_{r\in\mathcal{R}_g}\limits \! T_r \!\leq\!  \tau_g. \!\!\!\! &
  \vspace{1mm}
  \end{cases}
  \label{eq:CP}
\end{equation}
\normalsize


The first three constraints are related to the backhaul. The first guarantees that each relay obtains a minimum bandwidth, the second states that the aggregated bandwidth of relays is fixed, and the third is Shannon capacity.

The fourth constraint guarantees a minimum bandwidth for each served user, while the fifth and sixth constraints state that the aggregate share bandwidth of these users must adjust to the whole channel capacity allowed by their serving station. 

The seventh constraint restricts the throughput allocated to mobile users 
to the Shannon capacity.
The eighth constraint expresses the fact that the throughput allocated to relay-served users cannot exceed the wireless backhaul capacity assigned to the relay. Finally, the ninth constraint states that the aggregate throughput served by a \gNB (to mobile users \textsl{and} relays) cannot exceed the \gNB bottleneck over the wired backhaul.

The optimization program in~(\ref{eq:CP}) is 
convex, hence solvable in polynomial time with standard interior-point methods~\cite{nesterov1994interior}. Yet, such methods
have a cubic computational complexity with respect to the number of mobile users~\cite{bubeck2015convex}, which is prohibitive for real-time applications with large mobile user populations.
%
%
Thus, in the next section, we derive an exact analytical solution that has a linear complexity
with respect to the number of mobile users \textsl{and} relays attached to the \gNB.

\vspace{-2mm}
\section{The Exact $\max$--$\min$ Resource Allocation}
\label{a:OptimalCP}

In this section, we introduce \blue{\texttt{L\!i\!n\!E\!x}:} a scheme that provides\blue{, in \textsl{linear} time,} the \blue{\textsl{exact}} $\max$--$\min$ resource~allocation for the type of \blue{wireless relay} networks described in Section~II. 

\blue{The \texttt{L\!i\!n\!E\!x} scheme (cf. Algorithm~\ref{alg:BHmanaged})}
is independently executed at each \gNB $g$, and works as follows:

\begin{figure}[t]
\vspace{-0.25cm}
\begin{algorithm}[H]
  \caption{\blue{\texttt{L\!i\!n\!E\!x}: The \textsl{linear} and \textsl{exact} $\max$--$\min$ allocation.}
  }
  \label{alg:BHmanaged}
  \small
  \begin{algorithmic}[1]
  \vspace{-2.0mm}
    \STATE \textbf{Start:} \!\gNB~$g$, $w_{r} \!\leftarrow\! W_{\text{relays}}^{\min}$ and $T_r \!\leftarrow\! w_{r} \!\log_2\!\left(1 \!\!+\!\! \gamma_{g,r}\!\right)$, $\forall r\!\in\!\mathcal{R}_g$.\label{alg:BHmanagedStep2}
    \STATE {Derive optimal rates $\{T_u\}_{u\in\mathcal{U}_g \cup \; \mathcal{U}^*_g}$ for all users\blue{, limited to the wireless relay traffic of $T_r$ and ignoring the wired bottleneck.}\label{alg:BHmanagedStep3}}
    \STATE $\beta \leftarrow 1$. \label{alg:BHmanagedStep4}
    \WHILE{$\beta = 1$} \label{alg:BHmanagedStep5}
    \STATE $T_m \!\leftarrow\! \min \!\left\{T_{u} \!\!\suchthat\! T_{u} \!<\! w_u \log_2\!\left(1 \!+\! \gamma_{r, u}\right), r\!\in\!\mathcal{R}_g, u\!\in\!\mathcal{U}_r\!\right\}$.\label{alg:BHmanagedStep6}
    \STATE $\mathcal{L}_m \!\leftarrow\! \left\{u\!\in\! \mathcal{U}^{*}_g \suchthat T_{u} = T_m\right\}$. \label{alg:BHmanagedStep7}
    \STATE $T_M \!\leftarrow\! \min\left\{T_{u} \suchthat T_{u} > T_m,  u\!\in\!\mathcal{U}^{*}_g\right\}$.\label{alg:BHmanagedStep7.5}    
    \STATE $T_{M_2} \!\leftarrow\! \min\!\left(T_M\!, \min\!\left\{w_u \!\log_2\!\left(\!1 \!+\! \gamma_{r,u}\!\right) \!\!\suchthat\!\! r\!\in\!\mathcal{R}_g, u\!\in\!\mathcal{L}_m\!\cap \mathcal{U}_r \!\right\} \!\right)$.\vspace{1mm}\label{alg:BHmanagedStep8}
    \STATE $ \overline{\mathcal{U}}_r \!\leftarrow\! \left\{u\in\mathcal{U}_r\suchthat u\in\mathcal{L}_m\right\}$, $\forall r\in\mathcal{R}_g$. \label{alg:BHmanagedStep9}
    \STATE $\beta \!\leftarrow\! \min\left(1,\, \frac{W_\mathcal{\!\text{relays}}^g - \sum_{r\in\mathcal{R}_g}w_{r}}{\left(T_{M_2} - T_m\right)\cdot \sum_{r\in\mathcal{R}_g}|\overline{\mathcal{U}}_r|/\log_2\left(1+\gamma_{g,r}\right)}\right)$. \label{alg:BHmanagedStep10}
    \STATE $w_{r} \!\leftarrow\! w_{r} \!+\! |\overline{\mathcal{U}}_r| \beta \left(T_{M_2}\!-\!T_m\right)/\log_2\!\left(1\!+\!\gamma_{g,r}\right)$, $\forall r\!\in\!\mathcal{R}_g$.\label{alg:BHmanagedStep11}
    \STATE $T_r \!\leftarrow\! w_{r}  \log_2\left(1+\gamma_{g,r}\right)$, $\forall r\in\mathcal{R}_g$. \label{alg:BHmanagedStep12}
    \STATE $T_{u} \!\leftarrow\! T_{u} + \beta \left(T_{M_2}-T_m\right)$, $\forall u\in\mathcal{L}_m$. \label{alg:BHmanagedStep13}
    \ENDWHILE \label{alg:BHmanagedStep14}
    \STATE $T_r \!\leftarrow\! \sum_{u\in\mathcal{U}_r} T_{u}$, $\forall r\in\mathcal{R}_g$. \label{alg:BHmanagedStep15}
    \IF{$\sum_{u\in\mathcal{U}_g} \!T_u \!+\! \sum_{r\in\mathcal{R}_g} \!T_r>\tau_g$} \label{alg:BHmanagedStep17}
    \STATE{reduce the rates starting from the highest until the constraint on  $\tau_g$ in~(\ref{eq:CP}) is satisfied, preserving $\max$--$\min$ fairness.}
    \ENDIF \label{alg:BHmanagedStep18}
    \vspace{-0.5mm}
  \end{algorithmic}
  \normalsize
\end{algorithm}
\vspace{-8.0mm}
\end{figure}


\begin{figure*}
    \begin{center}
\vspace{-2mm}
\begin{tikzpicture}[-latex]
\hspace{-1.5mm}
  \matrix (chart)
    [
      matrix of nodes,
      column sep      = 1.4em,
      row sep         = 1.75ex,
    ]
    {
      |[root]| Network parameters setting and cell selection  &|[treenode]|  Assign minimum bandwidth and Shannon rate $T_r$ to relays & |[treenode]| Allocate user resources with 
      relay traffic $T_r$, ignoring the wired bottleneck & |[decision]| Is the minimum user rate $T_m$ lower than the Shannon capacity for all users with rate $T_m$? & |[treenode]| Raise the rate of all users with rate $T_m$ up to the minimum between their Shannon capacities and the next minimum 
      user rate
      \\
      & |[finish]| Final exact $\max$--$\min$ resource allocation&  |[treenode]| Reduce the excess of user rates in case of surpassing the wired backhaul traffic rate &|[decision2]| Have all users that had rate $T_m$ been able to increase their rate as much as possible?
      &|[treenode]| Raise relays' bandwidth and rate accordingly to afford the users' new rate\\
    };
  \draw
        (chart-1-1) edge (chart-1-2)
        (chart-1-2) edge (chart-1-3)
        (chart-1-3) edge (chart-1-4)
        (chart-1-4) \yes (chart-1-5)
        (chart-1-5) edge (chart-2-5)
        (chart-2-5) edge (chart-2-4)
        (chart-2-4) \yesR (chart-1-4)
        (chart-2-3) edge (chart-2-2);
  \draw
        (chart-2-4) \no (chart-2-3);
  \draw
        (chart-1-4) \noL +(-1.25,-1.15)
        (chart-1-4)+(-1.25,-1.15) -- (-2.05, 0) -- (chart-2-3);
\end{tikzpicture}
\vspace{-6mm}
\caption{\blue{Flowchart diagram of the \texttt{L\!i\!n\!E\!x} scheme operation.}}
\label{fig:flowchart}
\vspace{-6mm}
\end{center}
\end{figure*}

\begin{itemize}[leftmargin=\dimexpr\parindent+1.00mm+0.5\labelwidth\relax]
\setlength\itemsep{0.10cm}
\item
First of all, it assigns the minimum bandwidth $w_{r} \!=\! W_{\!\text{relays}}^{\min}$
and the highest achievable rate $T_r \!=\! w_{r} \log_2\!\left(1 \!+\! \gamma_{g, r}\right)$ to each relay $r\!\in\!\mathcal{R}_g$ (cf. Step~\ref{alg:BHmanagedStep2}).

\item
Then, it derives the optimal rates for all the users directly attached to either $g$ or to relays\blue{, limited to the relay~backhaul traffic of $T_r$ and ignoring the wired bottleneck}. Such subproblem has similarities with the one studied in~\cite{coluccia2012optimality}, whose solution is well known (\blue{for the interested reader, more details are provided in a separate technical report~\cite{arribas2023exact}, where} we show how this solution can be adapted to our system with one bottleneck).


\item
Now, we increase as much as possible the utilities by equally raising the lowest values of $\{T_{u}\}_{u\in\mathcal{U}_r}$, $\forall r\in\mathcal{R}_g$ (as long as constraints are not violated).
Let

\small
\vspace{-1mm}
\begin{eqnarray}
T_m \!=\! \min\limits_{u\in\mathcal{U}_r} \!\left\{T_{u} \!\!\suchthat\! T_{u} \!<\! w_u \log_2\!\left(1 \!+\! \gamma_{r, u}\right), r\!\in\!\mathcal{R}_g \right\}
\end{eqnarray}
\normalsize
be the minimum throughput rate that has not reached Shannon capacity (if $T_m$ does not exist, we are done). Let

\small
\vspace{-1mm}
\begin{eqnarray}
\mathcal{L}_m = \left\{u\!\in\! \mathcal{U}^{*}_g \suchthat T_{u} = T_m\right\}
\end{eqnarray}\normalsize
be the set of those relay--served users such that their rate is the same as the minimum $T_m$. Let

\small
\vspace{-1mm}
\begin{eqnarray}
T_M = \min\left\{T_{u} \suchthat T_{u} > T_m,  u\!\in\!\mathcal{U}^{*}_g\right\}
\end{eqnarray}\normalsize
be the minimum rate among relay--served user rates that are not as the minimum $T_m$ (cf. step~\ref{alg:BHmanagedStep7.5}).
Let\blue{'s further refine such minimum by considering the Shannon capacity of users in $\mathcal{L}_m$, which are in the worst serving condition:}

\small
\vspace{-0mm}
\begin{eqnarray}
T_{M_2} = \min\!\left(\!T_M, \min\limits_{r\in\mathcal{R}_g}\!\left\{w_u \!\log_2\!\left(1 \!+\! \gamma_{r,u}\!\right) \!\!\suchthat\!\! u\!\in\!\mathcal{L}_m\!\cap \mathcal{U}_r \right\} \!\right)\!.\!\! 
\end{eqnarray}\normalsize

The goal now is to increase $\{T_{u}\}_{u\in\mathcal{L}_m}$ as much as possible, without exceeding $T_{M_2}$, as long as those involved relays $r\in\mathcal{R}_g$ can request more resources to increase $T_r$. Let $\beta\in[0,1]$ be an auxiliary parameter that we will better define later. $\{T_{u}\}_{u\in\mathcal{L}_m}$ will be increased by $\beta (T_{M_2} - T_m)$, i.e., at most, by $T_{M_2} - T_m$ (cf. step~\ref{alg:BHmanagedStep13}). Let

\small
\vspace{-0mm}
\begin{eqnarray}\overline{\mathcal{U}}_r = \left\{u\in\mathcal{U}_r\suchthat u\in\mathcal{L}_m\right\}, \quad \forall  r\in\mathcal{R}_g \normalsize.
\end{eqnarray}\normalsize
Now, we set $T_{u}^{'} \!=\! T_{u} + \beta (T_{M_2}\!-\!T_m)$, $\forall u\in\mathcal{L}_m$~to~increase the involved throughput rates. Hence, we set $\forall r\!\in\!\mathcal{R}_g$:

\small
\vspace{-0mm}
\begin{eqnarray}
\!\!\!\!\!\!\!\!\!\!\!\!T_r \!\!&\!\!\!\!\!\!\!=\!\!\!\!& \!\!\!\!\!\!\sum\nolimits_{u\notin\overline{\mathcal{U}}_r}\!\!T_{u} \!+\!\sum\nolimits_{u\in\overline{\mathcal{U}}_r}\!\!T_{u}^{'} \nonumber\\[5pt] \!\!\!&\!\!\!=&\!\!\!\!\!\!\sum\nolimits_{u\notin\overline{\mathcal{U}}_r}\!\!T_{u} \!+\!\sum\nolimits_{u\in\overline{\mathcal{U}}_r}\!\!\!\left(T_{u} \!+\! \beta \left(T_{M_2}\!-\!T_m\right)\right) \nonumber\\[5pt] &\!\!\!\!\!\!=\!\!\!\!& \!\!\!\!\!\sum\nolimits_{u\notin\overline{\mathcal{U}}_r}\!\!T_{u} \!+\! \sum\nolimits_{u\in\overline{\mathcal{U}}_r}\!T_{u} + |\overline{\mathcal{U}}_r|\beta (T_{M_2}\!-\!T_m).
\label{eq:beta1}
\end{eqnarray}\normalsize
Hence, in step~\ref{alg:BHmanagedStep11} we set $\forall r\in\mathcal{R}_g$:

\small
\vspace{-0mm}
\begin{eqnarray}\!\!\!\!\!\!\!\!\!\!w_{r}^{new} &\!\!\!\!=\!\!\!\!& \frac{T_r}{\log_2\!\left(1\!+\!\gamma_{g,r}\!\right)}  = \frac{\sum_{u\in\mathcal{U}_r} \!\!T_{u} \!+\! |\overline{\mathcal{U}}_r|\beta (T_{M_2}\!\!-\!T_m\!)}{\log_2\!\left(1\!+\!\gamma_{g,r}\!\right)} \nonumber\\[5pt]
&\!\!\!\!=\!\!\!\!& w_{r} + \frac{|\overline{\mathcal{U}}_r|\beta (T_{M_2}\!-\!T_m)}{\log_2\left(1 + \gamma_{g,r}\right)}.
\end{eqnarray}\normalsize
The aggregation of the new relay resource allocation has to be lower than the total bandwidth, i.e.,

\small
\vspace{-0mm}
\begin{eqnarray}\!\!\!\!\!\!&&\!\!\!\sum\nolimits_{r\in\mathcal{R}_g} \!\!w_{r}^{n\!e\!w} = \!\sum\nolimits_{r\in\mathcal{R}_g}\!\left(w_{r} + \frac{|\overline{\mathcal{U}}_r|\beta\, (T_{M_2}\!-\!T_m)}{\log_2\!\left(1+\gamma_{g,r}\right)}\right) \nonumber\\[5pt]
\!\!\!\!\!\!&\!\!\!\!=\!\!\!\!& \!\!\!\sum\nolimits_{r\in\mathcal{R}_g} \!w_{r} \!+\! \beta (T_{M_2}\!-\!T_m)\!\sum\nolimits_{r\in\mathcal{R}_g} \!\frac{|\overline{\mathcal{U}}_r|}{\log_2\!\left(1\!+\!\gamma_{g,r}\!\right)}
\end{eqnarray} \normalsize
has to be lower than or equal to $W_{\!\text{relays}}^g$. Hence, isolating $\beta$ we get that necessarily:

\small
\vspace{-0mm}
\begin{eqnarray}
\beta \leq \frac{W_{\!\text{relays}}^g - \sum_{r\in\mathcal{R}_g} w_{r}}{(T_{M_2}-T_m) \sum_{r\in\mathcal{R}_g}|\overline{\mathcal{U}}_r|/\log_2\left(1+\gamma_{g,r}\right)}.
\end{eqnarray} \normalsize
Hence, in step~\ref{alg:BHmanagedStep10} we have defined $\beta$ as:

\small
\vspace{-0mm}
\begin{eqnarray}
\!\!\!\!\!\!\!\beta \!=\! \min\left(\!1, \frac{W_{\!\text{relays}}^g - \sum_{r\in\mathcal{R}_g} w_{r}}{(T_{M_2}\!\!-\!T_m) \sum_{r\in\mathcal{R}_g}\!|\overline{\mathcal{U}}_r|/\log_2\!\left(1\!+\!\gamma_{g,r}\right)}\right)\!.
\label{eq:beta5}
\end{eqnarray} \normalsize
Once the parameter $\beta$ is derived, we assign $w_{r} \!=\! w_{r}^{new}$ and $T_r \!=\! w_{r} \log_2\!\left(1 \!+\! \gamma_{g,r}\right)$, $\forall  r\in\mathcal{R}_g$ (cf. step~\ref{alg:BHmanagedStep12}).
In the case that $\beta = 1$ (cf. step~\ref{alg:BHmanagedStep5}), we repeat the process defining $T_m$ again and increasing the corresponding throughput~rates.

\item
To finalize the allocation and guarantee an exact solution, we need to ensure that the constraint on $\tau_g$ in~(\ref{eq:CP}) holds, which is done in steps~\ref{alg:BHmanagedStep17}--\ref{alg:BHmanagedStep18}. Whereas such reduction can be performed in a number of ways, in~\cite{arribas2023exact} we provide an algorithm preserves $\max$--$\min$ fairness: it reduces user throughputs from above $T\!=\!\min\{T_u\}$ down to $T$, at most, starting from the highest one, so that the aggregated network throughput reaches $\tau_g$; and if that is not enough, then assigns $T_u\!=\!\tau_g/|\mathcal{U}_g\!\cup\mathcal{U}_g^*|$.
\end{itemize}

\blue{For a better understanding of the \texttt{L\!i\!n\!E\!x} scheme, in Figure~\ref{fig:flowchart} we show a flowchart with a summary of the \texttt{L\!i\!n\!E\!x} operation.}


\vspace{-1mm}
\subsection*{\blue{Computational Complexity Analysis}}
\label{ss:complexity}

\blue{The \texttt{L\!i\!n\!E\!x} scheme guarantees the exact $\max$-$\min$ fairness. However, it is important to ensure that the proposed solution is deployable.}
Indeed,
\blue{the \texttt{L\!i\!n\!E\!x} scheme}
has a linear complexity in the number of the operations
with respect to the number of mobile users \textsl{and} the number of relays (i.e., the complexity is $\mathcal{O}\left(|\mathcal{U}_g\bigcup\mathcal{U}_g^*|\cdot |\mathcal{R}_g|\right)$, for each \gNB~$g$)\blue{, as shown next.}

In Algorithm~\ref{alg:BHmanaged},
\blue{the initial stage of deriving the user rates ignoring bottlenecks is solved in linear time with water-filling schemes~\cite{coluccia2012optimality}.}
\blue{Then,} 
the while loop will run over, at most, as many iterations as the number of relay-served mobile users. That happens because the while loop stops when $\beta\!<\!1$. However, that only happens when there are not enough resources to increase the resources for mobile users gathered in $\mathcal{L}_m$ (which grows, at least, by one mobile user at each iteration). Then, within the loop, we compute sums over the number of relays (i.e., $|\mathcal{R}_g|$)\blue{, as we thoroughly detail in \blue{a technical report}~\cite{arribas2023exact}.}
\blue{Afterwards, we sum the user rates for each relay and, finally, the excess of throughputs is optimally reduced to meet the wired bottleneck constraint with a linear descendent search.}
Hence, the overall complexity of \blue{the \texttt{L\!i\!n\!E\!x} scheme in} Algorithm~\ref{alg:BHmanaged} is $\mathcal{O}\!\left(|\mathcal{U}_g\bigcup\mathcal{U}_g^*|\!\cdot\! |\mathcal{R}_g|\right)$.

\section{\blue{Performance Evaluation}}
\blue{Here we present a performance evaluation of the \texttt{L\!i\!n\!E\!x} scheme. For that, we compare our proposal with two benchmarking schemes: the \texttt{CS\!o\!l\!v\!e\!r} and the \texttt{WF\!i\!l\!l} schemes.

On the one hand, \texttt{CS\!o\!l\!v\!e\!r} consists of a convex optimization \textsl{solver} that provides optimal solutions. Such optimizer has a high complexity (of the \textsl{cubic} order) that makes it undeployable in practice. However, it will allow us to verify that, indeed, our scheme provides optimal solutions.

On the other hand, the \texttt{WF\!i\!l\!l} scheme implements the solution of the $\max$--$\min$ resource allocation problem based on the known legacy allocation in~\cite{coluccia2012optimality}, following \textsl{water-filling} algorithms. Such a solution has been shown to be optimal when base stations~are considered individually, yet it does not take into account the~interwined nature of multiple-source allocations jointly constrained by wireless and wired bottlenecks. That is the main difference between the \texttt{WF\!i\!l\!l} and the \texttt{L\!i\!n\!E\!x} schemes: \blue{the former is the result of adapting the legacy scheduler to wireless relay networks, while} the latter \blue{has been thoughtfully designed to take into account} the backhaul resources and traffic constraints.}

\begin{figure}
\centering
\hspace{-0cm}
\vspace{-1mm}

\begin{minipage}[T]{.225\textwidth}
\hspace{-1.5mm}
\includegraphics[width=4.15cm]{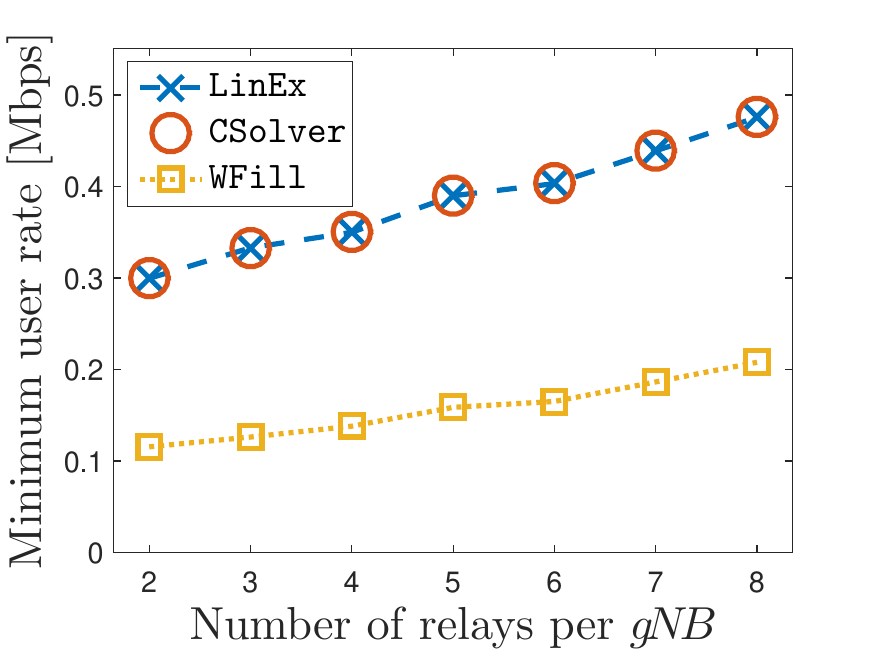}
    \vspace{-4.75mm}
    \caption{\blue{Wireless relay network with $3$ \gNBs and $U\!=\!600$ users.}}
\label{fig:VariationOfRelays}
\end{minipage}
%
\hspace{1mm}
%
\begin{minipage}[T]{.225\textwidth}
    \hspace{1mm}
    \includegraphics[width=4.15cm]{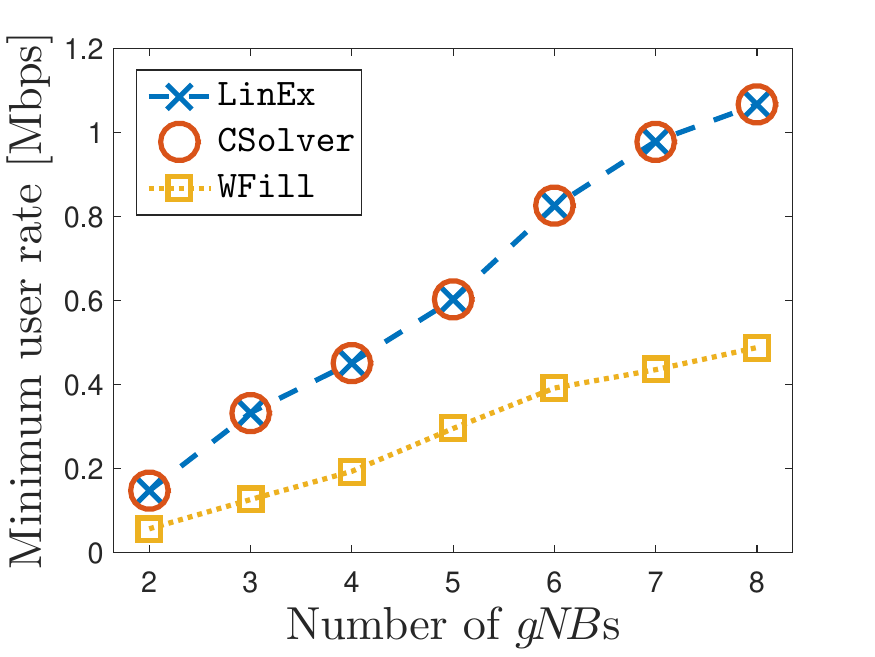}
    \vspace{-4.75mm}
    \caption{\blue{Wireless relay network with $3$ relays per \gNB and $U\!\!=\!600$.}}
\label{fig:VariationOfgNBs}
\end{minipage}
%
\vspace{-5mm} 
\end{figure}

\blue{All simulations are run over uniformly random network topologies in a circular region with radius of $750$~m.~Relays are considered as aerial relays, so that all network parameters and channel models are taken as in the realistic environment of~\cite{arribas2019coverage}: a heterogeneous dense urban network with terrestrial
path-loss~models and aerial line-of-sight (LoS)-based channel fading for relay-served users (for the interested reader, more details are provided in~\cite{arribas2023exact}).
The carrier frequency for \gNBs is 1815.1 MHz both for wireless backhaul (for transmissions to relays) and access channels (to mobile users), while for relays the carrier is 2630 MHz, with 20 MHz of band in all cases. Transmissions from \gNBs to relays do not interfere with transmissions from \gNBs to mobile users on the ground thanks to the adoption of precise 3D-beamforming over clear LoS links to the aerial relays.
Results are averaged over $1000$ runs.}

\blue{
In Figure~\ref{fig:VariationOfRelays} and Figure~\ref{fig:VariationOfgNBs} we observe the utility achieved (i.e., the minimum user rate) in two cases: $(i)$ when we increase the number of relays served by each \gNB in a network with $3$ \gNBs and $(ii)$ when we increase the number of \gNBs, each \gNB serving $3$ relays. In both cases there are $U=600$ mobile users that attach to the \gNB or relay cell with strongest signal (as in the operational 3GPP networks) and the wired bottleneck traffic is of $\tau_g = 180$~Mbps for each \gNB~$g$.}

\blue{Firstly, we see that the \texttt{L\!i\!n\!E\!x} and \texttt{CS\!o\!l\!v\!e\!r} schemes perform equally in all cases.
That means that \texttt{L\!i\!n\!E\!x} finds always the optimal $\max$--$\min$ resource allocation, with the important difference that \texttt{L\!i\!n\!E\!x} finds it in linear time, while the complexity of \texttt{CS\!o\!l\!v\!e\!r} is, instead, of the cubic order.
Secondly, we observe that as long as relays or \gNBs are added, network performance clearly increases. Indeed, the minimum user rate increases as users can find better connections and resource splitting opportunities.
Finally and most importantly, we remark that the performance of \texttt{WF\!i\!l\!l} is between $30\%$ to $60\%$ worse than \texttt{L\!i\!n\!E\!x}.
This shows that not only \texttt{L\!i\!n\!E\!x} is linear and exact, but it also considerably outperforms available state-of-the-art proposals.
Such a result reveals that it becomes crucial to account for the intertwined nature of multiple resource allocation at different cells, altogether constrained by backhaul resources and traffic rates.
Instead, simply adapting available allocation schemes to the wireless relay context is insufficient to achieve an acceptable network performance.
In conclusion, \texttt{L\!i\!n\!E\!x} stands as an efficient and lightweight implementable scheme for $\max$--$\min$ fair resource allocation in current wireless relay networks.}


\section{Conclusions}

We have solved the optimal $\max$--$\min$ allocation of downlink resources in wireless relay-enabled networks. With \blue{\texttt{L\!i\!n\!E\!x},} the \blue{proposed} exact $\max$--$\min$ \blue{resource} allocation \blue{scheme}, we have shown that the optimal distribution of resources can be found in linear time on the number of mobile users \textsl{and} 
relays, which is a key enabler for implementation over cellular networks.
Considering backhaul bottlenecks result to be crucial to assign resources to mobile users 
depending on the allocation to other relays and users.
We have shown that \blue{not only} our algorithm finds the optimal performance in terms of $\max$--$\min$ fairness \blue{in linear time, but it also stands as the only practical solution to enable $\max$--$\min$ fair resource allocation in wireless relay networks}.



\vspace{1mm}
\bibliographystyle{IEEEtran}
\bibliography{biblio}

\clearpage
\appendices
\newpage
\section{The Exact Allocation for Scenarios with One Base Station Without Relays}
\label{ss:BHfree}

In this scenario, we consider one base station~$s$ (namely, the \gNB or a relay) with no further relays attached to $s$. The CP
of Eq.~\eqref{eq:CP}
simplifies considerably since the station only needs to manage resources to be split among its served users $\mathcal{U}$: 

\small
\vspace{-2mm}
\begin{equation}
\hspace{-3.5cm}
  \begin{cases}
    \max \min \left\{T_u \suchthat u\in\mathcal{U}\right\}\!, & \text{s.t.:}\\
    w_u \geq W^{\min}, & \forall  u\!\in\!\mathcal{U}; \\
    \sum_{u\in\mathcal{U}} \! w_u = W_{\!\text{users}}; & \!\!\\
    T_u \!\leq\!  w_u \log_2\!\left(1 \!+\! \gamma_{s,u}\right)\!, & \forall u\!\in\!\mathcal{U}; \\
    \sum_{u\in\mathcal{U}} \! T_u \! \leq  \tau, &\!\!
  \end{cases}
  \label{eq:CPMaxMinEquiv}
\end{equation}
\vspace{-2mm}
\normalsize

\noindent
where $W_{\!\text{users}}$ and $W^{\min}$ are the total and minimum allocable bandwidth of the channel, while $\tau$ is the backhaul limitation.

The kind of strategies and algorithms to be followed in order to find the exact solution to such kind of subproblem with a single base station are well-known~\cite{coluccia2012optimality}. In particular, the exact solution adapted to our case is derived in Algorithm~\ref{alg:CPMaxMinEquiv}, followed by Algorithm~\ref{alg:MaxMinThrReduction} as a subroutine that will be eventually used to find the exact solution for the whole wireless relay network in Algorithm~\ref{alg:BHmanaged}. Algorithm~\ref{alg:CPMaxMinEquiv} is meant to find the optimal $\max$--$\min$ resource allocation when there is no capacity limitation (i.e., the $\tau$--constraint is ignored), while Algorithm~\ref{alg:MaxMinThrReduction} ensures that the $\max$--$\min$ rates are subsequently adjusted in a way that maintains the $\max$--$\min$ fairness found by Algorithm~\ref{alg:CPMaxMinEquiv} \textsl{and} the aggregate user rate does not exceed the backhaul capacity~limitation.

In Algorithm~\ref{alg:CPMaxMinEquiv}, we initially set $w_u\!=\!W^{\min}$ and $T_u \!=\! W^{\min}\!\cdot\! \log_2\!\left(1\!+\!\gamma_{s,u}\right)$, $\forall u\!\in\!\mathcal{U}$ (cf. step~\ref{alg:CPMaxMinEquiv:Step1}), and define the set $\mathcal{J}$ as those indices $u$ such that $T_u$ is minimum (cf. step~\ref{alg:CPMaxMinEquiv:Step3}):

\vspace{-0mm}
\begin{eqnarray}
  \mathcal{J}=\left\{u\in\mathcal{U}\suchthat T_u=\min\left\{ T_v \suchthat v\in \mathcal{U}\right\}\right\}.
  \label{eq:JPF}
\end{eqnarray}
\normalsize

While there are resources to allocate, i.e., $\sum_{u\in\mathcal{U}} \!w_u \!<\! W_{\!\text{users}}$, and $|\mathcal{J}|\!\!\neq\!\! |\mathcal{U}|$,
we take index $v_0\!=\!\arg\min_{u\notin\!\mathcal{J}}\!T_u$ so~that~$T_{v_0}$~is the lowest rate not equal to the minimum rate (cf. step~\ref{alg:CPMaxMinEquiv:Step5}).
Now, we aim to increase $w_u$~as~much~as possible in a way that is $\max$-$\min$ fair and $T_u\!\leq\! T_{v_0}\!,\forall u\!\!\in\!\!\mathcal{J}$. Then, we find $\{\!k_u\!\}_{u\!\in\!\mathcal{J}}$ so that $\{\!w_u\!\}_{u\!\in\!\mathcal{J}}$ are increased~by $k_u$ each. The optimal way is~by setting $k_u\!=\!\frac{T_{v_0}}{\log_2\!\left(\!1+\gamma_{s,u}\!\right)}\!-\!w_u$, $\!\forall u\!\!\in\!\!\mathcal{J}$ (cf. step~\ref{alg:CPMaxMinEquiv:Step6})~and~checking~if $\sum_{u\in\mathcal{J}}\!k_u\!\leq\! W_{\!\text{users}}\!-\!\sum_{u\in\mathcal{U}} \!w_u$. If that inequality is~not satisfied, then $k_u\!=\!\frac{W_{\!\text{users}}-\sum_{u\notin\mathcal{J}}w_u}{ \log_2\!\left(\!1+\gamma_{s,u}\!\right)\!\sum_{\!u\!\in\!\mathcal{J}}\!\!\frac{1}{ \log_2(\!1+\gamma_{s,u}\!)}}\!-\!w_u$, $\forall u\!\in\!\mathcal{J}$ (cf. step~\ref{alg:CPMaxMinEquiv:Step8}).

Once $k_u$ is derived, we assign $w_u \!\leftarrow\! w_u \!+\! k_u$, $\forall u\!\in\!\mathcal{J}$ (cf. step~\ref{alg:CPMaxMinEquiv:Step10}). Now, if we have set $k_u\!=\!\frac{T_{v_0}}{\log_2\!\left(1+\gamma_{s,u}\right)}-w_u$, $\forall u\!\in\!\mathcal{J}$, we re-set $\mathcal{J}$ as $\mathcal{J}\leftarrow \mathcal{J}\cup\{v_0\}$ (cf. step~\ref{alg:CPMaxMinEquiv:Step11}), and start all over. 

Note~\ref{note:CPMaxMinEquiv} shows that the $\{k_u\}_{u\in\mathcal{J}}$ of each iteration~of~Algorithm~\ref{alg:CPMaxMinEquiv} yields the optimal $\max$--$\min$ fair resource distribution.

\begin{note}
  Given a distribution of resources $\{w_u\}_{u\in\mathcal{U}}$ and throughput rates $\{T_u\}_{u\in\mathcal{U}}$ such that $T_u = w_u \log_2\left(1+\gamma_{s,u}\right)$, $\forall u\in \mathcal{U}$, we define the set $\mathcal{J}$ as in Eq.~\eqref{eq:JPF}.
  Hence, we have that $w_u \log_2\!\left(1\!+\!\gamma_{s,u}\right) = w_k \log_2\!\left(1\!+\!\gamma_{s,k}\right)$, $\forall u,k\in\mathcal{J}$.

  Given $v_0 = \arg\min\{T_u\suchthat u\notin\mathcal{J}\}$, we want to~increase $\{w_u\}_{u\in\mathcal{J}}$ as much as possible by $k_u$ each in~a $\max$--$\min$ fair way so that $(\!w_u\!\!+\!\!k_u\!)\!\log_2\!\left(\!1\!+\!\gamma_{s,u}\!\right) \!\leq\!  T_{\!v_0}$, $\forall u\!\in\!\mathcal{J}$.
  Hence, we must solve the following convex program:

\vspace{-0mm}
    \begin{equation}
    \hspace{-2cm}
    \begin{cases}
        \max L; \quad \text{s.t.:} & \\
        (w_u+k_u) \log_2\left(1+\gamma_{s,u}\right) \geq L, & \forall  u\!\in\!\mathcal{J}; \\
        k_u \leq  \frac{T_{v_0}}{\log_2\left(1+\gamma_{s,u}\right)} - w_u, & \forall u\!\in\!\mathcal{J}; \\
        \sum_{u\in\mathcal{J}} k_u \leq  W_{\!\text{users}}^g - \sum_{u\in\mathcal{U}} w_u. &
    \end{cases}
    \label{eq:CPMaxMinkiEquiv}
    \vspace{-0mm}
    \end{equation}
    \normalsize


   The KKT conditions~\cite{kuhn2014nonlinear} to solve the CP of Eq.~\eqref{eq:CPMaxMinkiEquiv}~are:

\vspace{-0mm}
    \small\begin{eqnarray}
    -\mu_u + \mu_{|\mathcal{J}|+u} + \mu_{2|\mathcal{J}|+1} = 0, & \forall  u\!\in\!\mathcal{J}; \label{eq:KKTCPMaxMinkiEquiv1}\\
    -1 + \sum_{u\in\mathcal{J}}\nolimits \mu_u/\log_2\left(1+\gamma_{s,u}\right) = 0; & \label{eq:KKTCPMaxMinkiEquiv2}\\
    \mu_u \cdot \left(L/\log_2(1\!+\!\gamma_{s,u})-k_u-w_u\right) = 0, & \forall u\!\in\!\mathcal{J}; \label{eq:KKTCPMaxMinkiEquiv3}\\
    \mu_{|\mathcal{J}|+u}\cdot  \left(k_u - T_{v_0}/\log_2(1\!+\!\gamma_{s,u})+w_u\right) = 0, & \forall u\!\in\!\mathcal{J}; \label{eq:KKTCPMaxMinkiEquiv4}\\
    \mu_{2|\mathcal{J}|+1} \!\cdot\!  \left(\sum_{u\in\mathcal{J}}\nolimits \! k_u - W_{\!\text{users}}^g - \sum_{u\in\mathcal{U}}\nolimits \!\! w_u \!\right) = 0. &\label{eq:KKTCPMaxMinkiEquiv5}
\vspace{-0mm}
  \end{eqnarray} \normalsize

If setting $k_u \!=\! T_{v_0}/\log_2\!\left(1\!+\!\gamma_{s,u}\right) \!-\! w_u$, $\forall u\!\in\!\mathcal{J}$~accomplishes that $\sum_{u\in\mathcal{J}} k_u \!\leq\!  W_{\!\text{users}}^g \!-\! \sum_{u\in\mathcal{U}} w_u$, we have the optimal solution, as each $k_u$ receives the maximum possible value and constraints hold.
Otherwise, $\exists u_0\!\in\!\mathcal{J} \!\!\suchthat\!\! k_{u_0} \!<\! T_{v_0}/\log_2\!\left(1\!+\!\gamma_{s,u_0}\!\right)\!-\!w_{u_0}$ and $\mu_{|\mathcal{J}|+u_0} \!=\! 0$ from Eq.~\eqref{eq:KKTCPMaxMinkiEquiv4}, and: 

\vspace{-0mm}
  \begin{equation}
    k_u = \frac{W_{\!\text{users}}^g-\sum_{u\notin\mathcal{J}}w_u}{\log_2\left(1+\gamma_{s,u}\!\right) \sum_{u\in\mathcal{J}} \frac{1}{\log_2(1+\gamma_{s,u})}}-w_u,  \forall  u\!\in\!\mathcal{J}.
  \end{equation}
  \normalsize
  \qed
  \label{note:CPMaxMinEquiv}
\end{note}

\vspace{-0mm}
If the resource allocation from Algorithm~\ref{alg:CPMaxMinEquiv} yields~a~feasible solution, i.e., $\sum_{u\in\mathcal{U}} \!T_u \!\leq \! \tau$, the optimal $\max$--$\min$ allocation is found. Otherwise,
due to the $\max$--$\min$ fairness nature, every mobile user~$u$ such that $T_u \!>\! \min_{u\in\mathcal{U}}\{T_u\}$ disposes of the minimum amount of~resources, $W^{\min}$, and $T_u = W^{\min}  \!\log_2\!\left(\!1\!+\!\gamma_{s,u}\!\right)$, $\forall u\!\in\!\mathcal{U} \!\suchthat\! T_u \!>\! \min_{u\in\mathcal{U}}\{T_u\}$ (otherwise,~if~such mobile users disposed of more than $W^{\min}$ resources, such exceeded resources could be reallocated to those mobile users with minimum rate to increase the utility, which is not possible from the $\max$--$\min$ fairness output).

\begin{figure}[t]
\vspace{-2.75mm}
\begin{algorithm}[H]
  \vspace{-2mm}
  \caption{Resource allocation without relays.}
  \label{alg:CPMaxMinEquiv}
  \small
  \begin{algorithmic}[1]
    \STATE \textbf{Start:} $\!w_u\!\leftarrow\! W^{\min}$, $T_u\!\leftarrow\! w_u \log_2\!\left(1\!+\!\gamma_{s,u}\!\right)$, $\!\forall u\!\in\! \mathcal{U}$.\label{alg:CPMaxMinEquiv:Step1}
    \STATE $\mathcal{J} \leftarrow  \left\{u\in\mathcal{U} \suchthat T_u = \min\left\{ T_v \suchthat v \in \mathcal{U}\right\}\right\}$.\label{alg:CPMaxMinEquiv:Step3}
    \WHILE{$\sum_{u\in\mathcal{U}} w_u < W_{\!\text{users}}$ \AND $|\mathcal{J}|\neq |\mathcal{U}|$}\label{alg:CPMaxMinEquiv:Step4}
    \STATE $v_0 \leftarrow \arg\min\{T_u\suchthat u\notin\mathcal{J}\}$ \AND $K \leftarrow 1$.\label{alg:CPMaxMinEquiv:Step5}
    \STATE $k_u \leftarrow T_{v_0}/\log_2\left(1+\gamma_{s,u}\right)-w_u$, $\forall u\in\mathcal{J}$.\label{alg:CPMaxMinEquiv:Step6}
    \IF{\!$\sum_{u\in\mathcal{J}}k_u > W_{\!\text{users}} - \sum_{u\in\mathcal{U}} w_u$}\label{alg:CPMaxMinEquiv:Step7} \STATE{$k_u \!\!\leftarrow\! \frac{W_{\!\text{users}}-\sum_{u\notin\mathcal{J}}w_u}{\log_2\!\left(\!1+\gamma_{s,u}\!\right)\!\sum\limits_{u\!\in\!\mathcal{J}} \!\!\frac{1}{\log_2\left(\!1+\gamma_{s,u}\!\right)}}\!-\!w_u$, $\!\forall u\!\in\!\mathcal{J}$ \AND $K \!\leftarrow\! 0$.}\label{alg:CPMaxMinEquiv:Step8} \ENDIF\label{alg:CPMaxMinEquiv:Step9}
    \STATE $w_u \!\!\leftarrow\! w_u \!+\! k_u$, $\!\forall u\!\in\!\mathcal{J}\!$ \AND $T_u \!\!\leftarrow\! w_u \!\log_2\!\left(\!1\!+\!\gamma_{s,u}\!\right)$, $\forall u\!\in\!\mathcal{J}$.\label{alg:CPMaxMinEquiv:Step10}
    \STATE{\textbf{if } $K = 1$, \textbf{ then } $\mathcal{J} \leftarrow \mathcal{J} \cup \{v_0\}$, \textbf{ end if}}\label{alg:CPMaxMinEquiv:Step11}
    \ENDWHILE\label{alg:CPMaxMinEquiv:Step14}
    \STATE \textbf{Output: } $T\leftarrow\min\left\{T_u \suchthat u\in \mathcal{U}\right\}$.\label{alg:CPMaxMinEquiv:Step15}
\vspace{-1mm}
  \end{algorithmic}
  \normalsize
\end{algorithm}
\vspace{-10mm}
\end{figure}

Hence, as no resources can be removed from any mobile user $u$ such that $T_u>\min_{u\in\mathcal{U}}\{T_u\}$, and $\sum_{u\in\mathcal{U}} T_u > \tau$, we must apply Algorithm~\ref{alg:MaxMinThrReduction} to the set $\mathcal{U}$ in order to reduce the rates in a way that keeps the $\max$--$\min$ fairness and accomplishes the $\tau$--constraint.
With that, the $\max$--$\min$ fairness allocation of the CP of Eq.~\eqref{eq:CPMaxMinEquiv} is finally solved.

\begin{note}
Algorithm~\ref{alg:MaxMinThrReduction} takes $\max$--$\min$ fair rates $\{T_u\}$ whose sum $A$ might exceed the traffic constraint $\tau$.
If that happens, it computes the initial $\max$--$\min$ fairness level with $T_{\min}$ as well as $(i)$ the excess aggregate throughput $E$ with respect to $\tau$, and $(ii)$ the aggregate surplus $S$, i.e., the sum of those rates in excess to $T_{\min}$ (cf. step~\ref{alg:CPMaxMinEquiv:StepIV}).
There are two cases. If the surplus is lower than the excess ($S \le E$), then eliminating the surplus will not be enough to meet the constraint on $\tau$. So, the only way is to assign each mobile user with equal resources $\tau/|\mathcal{U}|$ (cf. step~\ref{alg:CPMaxMinEquiv:Step17}) which, in turn, will be less (or at most as much as) the value $T_{\min}$. Otherwise, the surplus is more than the excess and the algorithm will reduce the surplus by exactly $E$, reducing the individual surplus of each user proportionally to its initial value (cf. step~\ref{alg:CPMaxMinEquiv:StepVIII}).
In all cases, the final rates are $\max$--$\min$ fair and do not exceed the capacity limitation~$\tau$.
\qed
\end{note}

\begin{figure}[t]
\vspace{-2.mm}
\begin{algorithm}[H]
  \caption{$\max$--$\min$ throughput reduction.}
  \label{alg:MaxMinThrReduction}
  \small
  \begin{algorithmic}[1]
  \vspace{-1mm}
    \STATE \textbf{Input:} Backhaul capacity limitation $\tau$, a set of users $\mathcal{U}$, and their $\max$--$\min$ fair rates $\{T_u\}_{u\in\mathcal{U}}$.
    \STATE $A = \sum_{u\in\mathcal{U}} \!T_u$.
    \IF{$A \!>\! \tau$}
    \label{alg:CPMaxMinEquiv:StepXII}
    \STATE
    $T_{\min}{=}\min\left\{T_u \suchthat u\in \mathcal{U}\right\}$,
    $E{=}A{-}\tau$,
    $S{=}\sum_{u\in\mathcal{U}} \left(T_u{-}T_{\min}\right)$.\label{alg:CPMaxMinEquiv:StepIV}
    \IF{$S \le E$}
    \label{alg:CPMaxMinEquiv:Step16}
    \STATE{$T_u \!\leftarrow\! \tau/|\mathcal{U}|$, $\forall u\!\in\!\mathcal{U}$.} \label{alg:CPMaxMinEquiv:Step17}
    \ELSE
    \STATE{$T_u \!\leftarrow\! T_u - E \cdot \frac{T_u-T_{\min}}{S}$, $\forall u\!\in\!\mathcal{U}$.}\label{alg:CPMaxMinEquiv:StepVIII}
    \ENDIF
    \ENDIF\label{alg:CPMaxMinEquiv:StepXVII}
    \STATE \textbf{Output: } $\{T_u\}_{u\in\mathcal{U}}$. \label{alg:CPMaxMinEquiv:Step15}
    \vspace{-0.mm}
  \end{algorithmic}
  \normalsize
\end{algorithm}
\vspace{-6.5mm}
\end{figure}

\section{Linear Complexity of the Exact Resource Allocation of Algorithm~\ref{alg:BHmanaged}}
\label{app:s:complexity}
The
exact $\max$--$\min$ resource allocation provided by Algorithm~\ref{alg:BHmanaged} has a linear complexity in the number of the operations with respect to the number of mobile users \textsl{and} the number of relays (i.e., $\mathcal{O}\!\left(|\mathcal{U}_g\bigcup\mathcal{U}_g^*|\cdot |\mathcal{R}_g|\right)$, for each~$g$). Here, we provide the details of that result.

As
Algorithm~\ref{alg:BHmanaged} runs Algorithms~\ref{alg:CPMaxMinEquiv} and~\ref{alg:MaxMinThrReduction} as subroutines, we analyze the complexity of both algorithms first and integrate their complexity later to the full complexity of Algorithm~\ref{alg:BHmanaged}.

\subsection{Algorithm~\ref{alg:CPMaxMinEquiv}: Resource allocation of one base station without~relays}
The
complexity of Algorithm~\ref{alg:CPMaxMinEquiv} is tricky because, apparently, it could seem that it has a quadratic complexity.
However, we show
that, instead, it has a linear complexity of $\mathcal{O}\!\left(| \mathcal{U} |\right)$.

Basically,
on the one hand, the number of loop iterations is, at most, the number of users. On the other hand, the sums within the loop can be easily rearranged to be derived with a fixed number of operations, based on what has been computed in the previous iteration.
Also,
it can be assumed, without loss of generality, that the unit Shannon capacities of users (i.e., $\{\log(1\!+\!\gamma_{s,u})\}$) are sorted in increasing order, so that deriving minimums becomes immediate due to sorting.

More formally, we analyze Algorithm~\ref{alg:CPMaxMinEquiv} step by step and show that it has indeed a linear complexity of $\mathcal{O}\!\left(| \mathcal{U} |\right)$.

First, in step~\ref{alg:CPMaxMinEquiv:Step1} we can assume, without loss of generality, that the unit Shannon capacities of the input users (i.e., $\{\log(1\!+\!\gamma_{s,u})\}$) are sorted in increasing order. That is because any base station running Algorithm~\ref{alg:CPMaxMinEquiv} receives the CSI of each user and can sort the SINRs $\gamma_{s,u}$ along. Next, in step~\ref{alg:CPMaxMinEquiv:Step3} we compute the set $\mathcal{J}$ for the first time as those users with minimum rate.

Now,
the while loop
has, at most, as many iterations as
the number of
mobile users,
since the loop iterates while
the set $\mathcal{J}$ still has users to gather. But, in step~\ref{alg:CPMaxMinEquiv:Step11}, we see that $\mathcal{J}$ gets exactly one
user
per
iteration, although only when $K=1$. But
$K$
is
equal to $1$ only when the condition of step~\ref{alg:CPMaxMinEquiv:Step7} holds, which is equivalent to the condition of the while loop. Hence, if $K=0$, the loop~stops.

From the next step on we see that, apparently, at each iteration the algorithm must compute some sums over a set of users. However, those sums can be easily rearranged to be derived with a fixed number of operations, on top of the sums computed in the previous iteration. To see that, we will need to know, given some set $\mathcal{J}$, the value $S_{|\mathcal{J}|}$, defined as:

\small
\begin{equation}
  S_{|\mathcal{J}|} = \sum_{u\in\mathcal{J}}\nolimits \frac{1}{\log_2(1+\gamma_{s,u})}.
  \label{eq:SJ}
\end{equation}
\normalsize

To compute $S_{|\mathcal{J}|}$, we note that there are only $|\mathcal{U}|$ possible sets $\mathcal{J}$ in the execution of Algorithm~\ref{alg:CPMaxMinEquiv}. Indeed, set $\mathcal{J}$~is composed by those users with minimum rate and then the algorithm adds exactly one user per iteration to $\mathcal{J}$, which is to the next user with minimum rate (cf. step~\ref{alg:CPMaxMinEquiv:Step5}). Since rates $T_u$ are not modified until they are included in $\mathcal{J}$, we can know before the while loop what are the $|\mathcal{U}|$ possible sets $\mathcal{J}$, all with different cardinalities, that Algorithm~\ref{alg:CPMaxMinEquiv} could use.

Concretely, if we first consider the smallest possible set $\mathcal{J}$ with one user, we can compute $S_{|\mathcal{J}|}$ as:

\small
\vspace{-0mm}
\begin{eqnarray}
  S_1 = \frac{1}{\log_2(1+\gamma_{s,1})}.
\end{eqnarray}
\normalsize

Now, if instead we consider some possible set $\mathcal{J}$ with more than one user, it can be readily seen, from Eq.~\eqref{eq:SJ}, that:

\small
\vspace{-0mm}
\begin{eqnarray}
  S_{|\mathcal{J}|} = S_{|\mathcal{J}|-1} + \frac{1}{\log_2(1+\gamma_{s,|\mathcal{J}|})}.
\end{eqnarray}
\normalsize

Hence, to compute $S_{|\mathcal{J}|}$, we only need the sum of the~previous term $S_{|\mathcal{J}|-1}$ with the term $\frac{1}{\log_2(1+\gamma_{s,|\mathcal{J}|})}$. Since there~are, at most, $|\mathcal{U}|$ possible sets $\mathcal{J}$, we can compute all possible values of $S_{|\mathcal{J}|}$ with only $|\mathcal{U}|$ sums, before running the loop. 


Now, we can see that within the while loop we need to know the value of $\sum_{u\in\mathcal{U}} w_u$, which depends on the new $w_u$ values that have been derived at the end of the previous iteration in step~\ref{alg:CPMaxMinEquiv:Step10}. But that sum is equal to:

\small
\vspace{-0mm}
\begin{eqnarray}
  \sum_{u\in\mathcal{U}}\nolimits w_u = \sum_{u\in\mathcal{J}}\nolimits w_u + \sum_{u\notin\mathcal{J}}\nolimits w_u.
\end{eqnarray}
\normalsize

Now, we can see that, on the one side, users in $\mathcal{J}$ have been computed in the previous iteration as $w_u\leftarrow w_u^{old} + k_u$ in step~\ref{alg:CPMaxMinEquiv:Step10} and, checking the value assigned to $k_u$ in step~\ref{alg:CPMaxMinEquiv:Step6}, we have that, in reality, $w_u=T_{v_0}/\log_2(1+\gamma_{s,u})$, for those users $u\in\mathcal{J}$. On the other side, the $w_u$ value of those users not included in $\mathcal{J}$ is equal to $w_u=W^{\min}$, because those users have not been selected yet to be included in $\mathcal{J}$.
Hence:

\small
\vspace{-0mm}
\begin{eqnarray}
  \!\!\!\!\!\!\!\!\sum_{u\in\mathcal{U}}\nolimits w_u &\!\!\!=\!\!\!&\!\sum_{u\in\mathcal{J}}\nolimits w_u + \sum_{u\notin\mathcal{J}}\nolimits w_u \nonumber\\ &\!\!\!=\!\!\!&\!T_{v_0}\!\sum_{u\in\mathcal{J}}\nolimits\frac{1}{\log_2(1\!+\!\gamma_{s,u})} \!+\! (|\mathcal{U}| \!-\! |\mathcal{J}|)W^{\min}.
  \label{eq:sumSJ}
\end{eqnarray}
\normalsize

Now, we note that the only sum remaining in Eq.~\eqref{eq:sumSJ} is $\sum_{u\in\mathcal{J}} \frac{1}{\log_2(1+\gamma_{s,u})}$, which is the value of $S_{|\mathcal{J}|}$ computed in advance, before running the loop. Hence, there is no need to do $|\mathcal{J}|$ sums, as  $S_{|\mathcal{J}|}$ is already known.
Thus, computing $\sum_{u\in\mathcal{U}} w_u$ can be done with a fixed number of operations.

Regarding step~\ref{alg:CPMaxMinEquiv:Step5}, we note that it incurs no cost because rates $T_u$ not included in $\mathcal{J}$ are still sorted since step~\ref{alg:CPMaxMinEquiv:Step1}.

Then, we can rearrange step~\ref{alg:CPMaxMinEquiv:Step6}: Let's define, for each $u\!\in\!\mathcal{J}$, the variable $L_u$ (variable $L_u$ would be equivalent to $L_u\!=\!(k_u\!+\!w_u)\log_2(1\!+\!\gamma_{s,u})$) and change step~\ref{alg:CPMaxMinEquiv:Step6} by the following~line:

\small
\vspace{-0mm}
\begin{eqnarray}
  L_u\leftarrow T_{v_0}, \quad \forall u \in \mathcal{J}.
\end{eqnarray}
\normalsize

Now, step~\ref{alg:CPMaxMinEquiv:Step6} is the same but with a change of variables. With that, we have $|\mathcal{J}|$ variable assignments with no operations.

Regarding step~\ref{alg:CPMaxMinEquiv:Step7}, we need to check whether $\sum_{u\in\mathcal{J}} k_u \!+\!\sum_{u\in\mathcal{U}} w_u \!>\! W_{\text{users}}$. Thus, since we know that here $k_u\!=\!T_{v_0}/\log_2(1\!+\!\gamma_{s,u})-w_u$, then:

\small
\vspace{1mm}
\small
\begin{eqnarray}
  \!\!\!\!\!\!\!\!\!\!\!&\!\!\!\!\!&\sum_{u\in\mathcal{J}}k_u \!+\! \sum_{u\in\mathcal{U}} w_u =\sum_{u\in\mathcal{J}}\!\!\left(\!\frac{T_{v_0}}{\log_2(1\!+\!\gamma_{s,u})} \!-\! w_u\!\!\right) \!+\! \sum_{u\in\mathcal{U}} \!w_u  \nonumber\\
  \!\!\!\!\!\!\!\!\!\!\!\!\!&=\!\!\!\!\!&T_{v_0}\sum_{u\in\mathcal{J}}\nolimits\frac{1}{\log_2(1\!+\!\gamma_{s,u})} - \sum_{u\in\mathcal{J}}\nolimits w_u + \sum_{u\in\mathcal{U}}\nolimits w_u \nonumber\\
  \!\!\!\!\!\!\!\!\!\!\!\!\!&=\!\!\!\!\!&T_{v_0}\sum_{u\in\mathcal{J}}\nolimits\frac{1}{\log_2(1\!+\!\gamma_{s,u})} + \sum_{u\notin\mathcal{J}}\nolimits w_u \nonumber\\ \!\!\!\!\!\!\!\!\!\!\!\!\!&=\!\!\!\!\!&T_{v_0}\sum_{u\in\mathcal{J}}\nolimits\frac{1}{\log_2(1\!+\!\gamma_{s,u})} + \sum_{u\notin\mathcal{J}}\nolimits W^{\min}\nonumber\\
   \!\!\!\!\!\!\!\!\!\!\!\!\!&=\!\!\!\!\!&T_{v_0}\sum_{u\in\mathcal{J}}\nolimits\frac{1}{\log_2(1\!+\!\gamma_{s,u})} + (|\mathcal{U}| - |\mathcal{J}|)W^{\min}. \label{eq:last}
\end{eqnarray}
\normalsize

\vspace{1mm}
So, we see that the sum over the users in $\mathcal{J}$ that appears in Eq.~\eqref{eq:last} is, indeed, the value $S_{|\mathcal{J}|}$ computed in advance, so that it needs no extra operations.

For step~\ref{alg:CPMaxMinEquiv:Step8} we can simply observe that this step will be run, at most, once. That is because if we get to this step, that means that the condition of step~\ref{alg:CPMaxMinEquiv:Step7} holds, which in turn means that the condition of the while loop in step~\ref{alg:CPMaxMinEquiv:Step4} does not hold anymore and the algorithm stops.

Step~\ref{alg:CPMaxMinEquiv:Step10} can be avoided and leave the final assignment of those values $w_u$ and $T_u$ for when the while loop has finished. The important thing here is, according to how we have rearranged some steps, how the new variable $L_u$ varies at each iteration. Indeed, in order to derive $w_u$ and $T_u$ of users in $\mathcal{J}$ once the while loop has finished (since $w_u$ and $T_u$ of users not in $\mathcal{J}$ have not changed), we can simply set:

\small
\vspace{-0mm}
\begin{equation}
  w_u = \frac{L_u}{\log_2(1+\gamma_{s,u})}, \text{ and } T_u = L_u, \quad \forall u\!\in\!\mathcal{J}.
\end{equation}
\normalsize

As a result, Algorithm~\ref{alg:CPMaxMinEquiv} does have a linear complexity with respecto to the number of users $|\mathcal{U}|$, i.e., $\mathcal{O}\left(|\mathcal{U}|\right)$.

\subsection{Algorithm~\ref{alg:MaxMinThrReduction}: $\max$--$\min$ throughput reduction}
Algorithm~\ref{alg:MaxMinThrReduction} is clearly linear with respect to the number of mobile users $|\mathcal{U}|$ because there are no loops and there are just a couple of sums over the number of mobile users. Moreover, deriving the minimum of step~\ref{alg:CPMaxMinEquiv:StepIV} would take at most $|\mathcal{U}|$ comparisons, even ignoring that rates could be sorted.

As a result, Algorithm~\ref{alg:MaxMinThrReduction} has a linear complexity of $\mathcal{O}\!\left(|\mathcal{U}|\right)$.

\subsection{Overall complexity of Algorithm~\ref{alg:BHmanaged}: The exact $\max$--$\min$ resource allocation}

Now, let us focus on Algorithm~\ref{alg:BHmanaged}. Initially, Algorithm~\ref{alg:CPMaxMinEquiv} is run as a subroutine once for the \gNB~$g$ and then for each relay $r$ in $\mathcal{R}_g$. Since we have seen that Algorithm~\ref{alg:CPMaxMinEquiv} has linear complexity with respect to the number of users, at this point we already have a complexity of at most $\mathcal{O}\!\left(|\mathcal{U}_g\bigcup\mathcal{U}_g^*|\!\cdot\! |\mathcal{R}_g|\right)$.

After that, basically, it can be seen that the while loop will have, at most, as many iterations as the number of relay-served users. That happens because the while loop stops when $\beta\!<\!1$. However, that only happens  when there are not enough resources to increase the resources for links gathered in $\mathcal{L}_m$ (which grows, at least, by one link at each iteration). Then, within the loop, we compute either a fixed number of sums over the number of relays in $\mathcal{R}_g$ or a fixed number of operations for each relay in $\mathcal{R}_g$.
Hence, the overall complexity of~Algorithm~\ref{alg:BHmanaged} is of the order of $\mathcal{O}\!\left(|\mathcal{U}_g\bigcup\mathcal{U}_g^*|\!\cdot\! |\mathcal{R}_g|\right)$, in the worst case.

More formally, let's see that the number of iterations of Algorithm~\ref{alg:BHmanaged} is lower than the number of users so that the overall complexity is linear with respecto to both, the number of users and the number of relays.

Clearly, the while loop ends when $\beta < 1$. The goal of that loop is to take always the lowest relay--served user rate $T_m = T_{u}$, $u\in\mathcal{U}_g^{*}$ and rise that rate as much as possible. In the case that there were more users with the same minimum rate, the algorithm rises them all simultaneously, in order to be $\max$--$\min$ fair.
Such users are the ones contained in the set $\mathcal{L}_m$. Now, the algorithm rises all their rates up to the next lowest rate (but higher than their own rate), $T_M$, taking into account that in the case that the Shannon capacity of some user would not be enough to reach $T_M$, then the algorithm would rise the rates up to the lowest Shannon capacity of that user, i.e., $T_{M_2}$. Once it is not possible to reach $T_{M_2}$, the algorithm should stop because there would be minimum rates that could not be risen any more so that the state would already be $\max$--$\min$ fair.

For those chosen minimum relay-served user rates $T_{u}$ (all with value equal to $T_m$) that have to be risen up to, if possible, $T_{M_2}$, we would need to add up the amount of $(T_{M_2} - T_m)$, since $T_m + (T_{M_2}-T_m) = T_{M_2}$. But it is important to notice that we would be rising the rates without having into account whether the backhaul link of relay $r$ has still the possibility to take resources that have not been assigned yet. Hence, we do not rise the rates by $(T_{M_2}-T_m)$ but by $\beta(T_{M_2}-T_m)$ instead, where $0\leq \beta\leq 1$. Then, the value $\beta$ is computed in Eqs.~\eqref{eq:beta1}--\eqref{eq:beta5} in order to make sure that the amount of assigned backhaul resources is not higher than the actual amount of available backhaul resources of relays. Hence, in the case that there are indeed enough available backhaul resources for relays, $\beta$ will be equal to $1$ and the while loop will keep iterating. Otherwise, having $\beta<1$ will mean that there were not enough backhaul resources available to rise the rates as much as planned, i.e., value $T_{M_2}$ will not be reached by the risen rates, and the $\max$--$\min$ state will have been reached and the while loop will end.

Therefore, as within the while loop the algorithm always takes all users with minimum rate at the beginning of each iteration, in the set $\mathcal{L}_m$ we accumulate those relay--served users with lowest rate. Then, that set contains always the same users and, at each iteration, it adds at least one new user. Thus, there will be, at most, as many iterations as the number of relay--served users, $|\mathcal{U}_g^{*}|$.

Now, within the while loop the number of operations are as follows. The computation of the minimums are immediate if rates are sorted in advance in a sub-quadratic number of comparisons, which are much faster compared to actual operations. Hence, what matters is that in step~\ref{alg:BHmanagedStep10} the algorithm computes a couple of sums over the number of relays, i.e., $|\mathcal{R}_g|$, and then a fixed number of operations, in order to compute the value to be assigned to $\beta$. Then, in steps~\ref{alg:BHmanagedStep11} and~\ref{alg:BHmanagedStep12} a fixed number of operations is computed for each relay in $\mathcal{R}_g$.
Next in step~\ref{alg:BHmanagedStep13}, the assigned value to each $T_{u}$ is the same for all those users in $\mathcal{L}_m$, so that the fixed number of operations of that step are computed only once, and then assigned to each user.

Finally, once the while loop ends, we compute in steps~\ref{alg:BHmanagedStep15} and~\ref{alg:BHmanagedStep17} some sums over the number of mobile users or the number of relays, which need as many sums as either the number of mobile users or the number of relays. Next, Algorithm~\ref{alg:MaxMinThrReduction} is maybe applied, which we already know that has linear complexity.

As a result, integrating the complexity of the subroutines of Algorithms~\ref{alg:CPMaxMinEquiv} and~\ref{alg:MaxMinThrReduction}, the overall complexity of Algorithm~\ref{alg:BHmanaged} is linear with respect to both, the number of mobile users \textsl{and} the number of relays, i.e., $\mathcal{O}\!\left(|\mathcal{U}_g\bigcup\mathcal{U}_g^*|\cdot |\mathcal{R}_g|\right)$.

\balance

\section{Channel Models and Parameters Used in the Performance Evaluation}
\label{ss:parameters}

All simulations are run over uniformly random network topologies in a circular region with radius of $750$~m.~Relays are considered as aerial relays or, equivalently, aerial base stations (\aBS), so that all network parameters and channel models are taken as in the realistic environment of~\cite{arribas2019coverage}: a heterogeneous dense urban network with terrestrial
conventional
path-loss~models and aerial line-of-sight (LoS)-based channel fading for relay-served users.
The carrier frequency for \gNBs is 1815.1 MHz both for wireless backhaul (for transmissions to relays) and access channels (to mobile users), while for relays the carrier is 2630 MHz, with 20 MHz of band in all cases. Transmissions from \gNBs to relays do not interfere with transmissions from \gNBs to mobile users on the ground thanks to the adoption of precise 3D-beamforming over clear LoS links to the aerial relays.
Results are averaged over $1000$ runs.

In Table~\ref{t:parameters} we report the evaluation parameters used in our numerical results.

\begin{table}[H]
\caption{Evaluation parameters}
\vspace{-1mm}
    \label{Parameters}
    \centering
    \begin{tabular}{|c|c|} 
    \hline
       \textbf{\textsl{Parameter}} & \textbf{\textsl{Value} }  \\ \hline \hline
       $\xi_{LoS}$, $\xi_{NLoS}$, $\beta_1$, $\beta_2$ & 1.6 dB, 23 dB, 12.08, 0.11 \\ \hline
       Carrier frequencies, $f_\mathcal{G}$, $f_{\!\mathcal{A}}$ & $1815.1$ MHz, $2.63$ GHz \\ \hline
       Bandwidths, $W_\mathcal{G}$, $W_{\!\mathcal{A}}$& $20$ MHz, $20$ MHz  \\ \hline
       Tx power, $P_{Tx}^g$, $P_{Tx}^a$ & $44$ dBm, $25$ dBm \\ \hline
       Thermal Noise Power & -174 dBm/Hz  \\ \hline
       Ground path loss exponent, $\eta_{\mathcal{G}}$ & $3$ \\ \hline
       Height range, $[h_{\min}, h_{\max}]$ & $[40, 300]$ m \\ \hline
       Instances of simulations & $1000$ \\ \hline
    \end{tabular}
    \vspace{0mm}
    \label{t:parameters}
\end{table}

In what follows, we provide the details of the channel modelling followed in the network.

\subsection{Channel Modelling}
\label{a:ss:pathloss}

We assume that the network operator disposes of two orthogonal frequency bands. One band is assigned to \gNBs to provide access service to ground users as well as aerial backhaul service to \aBSs. 
The other band is assigned to \aBSs for aerial user access.
Hence, we model three different channel types: $(i)$ air-to-ground and $(ii)$ ground-to-ground channels in the access network, and $(iii)$ ground-to-air channels in the backhaul network.

Indeed, the access network communication channels between serving base stations and UEs differ much depending on whether users connect to a \gNB or to an \aBS.
While the ground-to-ground channel attenuation for \gNB--UE links follows conventional path-loss modeling based on slow and fast fading, air-to-ground channels (\aBS--UE links) suffer additional attenuation depending on the LoS---or NLoS---state of the channel.
Such additional attenuation is referred to in the literature as an \textsl{excess attenuation}~\cite{al2014modeling}. Moreover, antennas used for the access network differ from backhaul network antennas performing 3D-beamforming, which directly affects the interference suffered in each case. In the following sections, we detail these features for each type of modelled channel.

\subsubsection{Air-to-Ground Channels}
\label{a:ss:A2G}

Depending on whether links between \aBSs and UEs are free of obstacles (e.g., buildings, traffic, etc.), the attenuation differs notably~\cite{al2014modeling}.
The LoS-likelihood is a complex function of the elevation angle between UE $u\in\mathcal{U}$ and \aBS $a\in\mathcal{A}$:

\small
\begin{eqnarray}
    P_{LoS}(a, u) = \frac{1}{1+\beta_1\cdot \exp\!\left(-\beta_2\left(\frac{180}{\pi}\arctan\left(\frac{h_a}{r_{a,u}}\right)-\beta_1\right)\right)},
\label{eq:LoS}
\end{eqnarray}
\normalsize
where the elevation of $a$ is $h_a$, while $\beta_1$ and $\beta_2$ are parameters depending on the number of large signal obstructions per unit area, building's height distribution, ratio of built-up area and clean surfaces, etc., as derived in~\cite{al2014optimal}, based on ITU recommendations~\cite{recommendations2015propagation}.
In Eq.~\eqref{eq:LoS}, $\theta_{a, u}\!=\!\arctan(h_a / r_{a, u})$ is the elevation angle.
$\theta_{a, u}$ approaches $\frac{\pi}{2}$ when the \aBS~$a$ hovers just above the user~$u$, i.e., when the LoS likelihood reaches its maximum.
The elevation angle $\theta_{a, u}$ is characterized by the \aBS height and the ground distance between the user and the \aBS, that is
$r_{a,u}$.
%

In particular, the average attenuation (in dB units) of an air-to-ground channel between drone $a$ and user $u$ depends on the LoS likelihood, with the following expression~\cite{al2014optimal}:

\small
\begin{eqnarray}
    L_\mathcal{A}(a,u) \!\!&=&\!\! 20\log_{10}\left(\frac{4\pi f_{\!\mathcal{A}}}{c} \cdot \sqrt{h_a^2 + r_{a,u}^2} \right) + \\ && P_{LoS}(a, u) \cdot \left(\xi_{LoS} - \xi_{N\!LoS}\right) + \xi_{N\!LoS},
    \label{eq:PLdrone}
\end{eqnarray}
\normalsize
where $\xi_{LoS}, \xi_{N\!LoS}$ are the {\it excess attenuation} components in LoS/NLoS conditions; $f_{\!\mathcal{A}}$ is the carrier frequency in~Hz; and $c$ is the speed of light in~m/s.

Since \gNBs and \aBSs operate onto orthogonal bands, there is no interference between drone-served users and cellular users.
With the above, the experienced SINR for air-to-ground access links $(a, u)$ is:

\small
\begin{eqnarray}
  \gamma^\mathcal{A}_{a, u} = \frac{P^a_{Tx} \!\cdot\! 10^{-L_\mathcal{A}(a, u)/10}}{N_{a, u} + I_{a, u}^\mathcal{A}},
\end{eqnarray}
\normalsize

\noindent
where $P^a_{Tx}$ is the transmission power of an omnidirectional antenna in the \aBSs $a\in\mathcal{A}$;
$N_{a, u}$ is thermal noise according to the allocated bandwidth; and $I_{a,u}^\mathcal{A}$ is the interference level that user~$u$ suffers from other \aBSs. However, note that the 3D position of an \aBS is a decision parameter that directly affects interfering signals received by user $u$, i.e.:

\small
\begin{eqnarray}
  I_{a, u}^\mathcal{A} = \sum\limits_{a'\in\mathcal{A}\setminus\{a\}} P^a_{Tx} \!\cdot\! 10^{-L_\mathcal{A}(a', u)/10}, \quad \forall a\in\mathcal{A},
\end{eqnarray}
\normalsize

\noindent
where $L_\mathcal{A}(a', u)$ depends on the 3D position of \aBSs $a'\!\in\!\mathcal{A}$, as shown in Eq.~\eqref{eq:PLdrone}.

\subsubsection{Ground-to-Ground Channels.}
\label{a:ss:G2G}
Connections in the access network between \gNBs and users experience an attenuation based on a \blue{well-known} path-loss model with slow fading (in dB units):

\small
\begin{eqnarray}
    L_\mathcal{G}(g, u) = 10\eta_\mathcal{G}\log_{10}\left(\frac{4\pi f_\mathcal{G}}{c_{l}} \cdot  \text{dist}(g,u)\right)  + \mathcal{N}(0, \sigma_\mathcal{G}^2),
  \label{eq:PLeNB}
\end{eqnarray}
\normalsize

\noindent
where $\eta_\mathcal{G}>2$ is the path-loss exponent in ground communications; $f_\mathcal{G}$ is the operating carrier frequency of the \gNBs; and $\sigma_{\mathcal{G}}$ is the standard deviation of the \blue{Gaussian} random variable $\mathcal{N}(0, \sigma^2_\mathcal{G})$, modelling the effects of shadowing.

As mentioned above, since there is no interference between cellular users and drone-served users, the SINR for access links $(g, u)$ is:

\small
\begin{eqnarray}
  \gamma^\mathcal{G}_{g, u} = \frac{P^g_{Tx} \!\cdot\! 10^{-L_\mathcal{G}(g, u)/10}}{N_{g, u} + I_{g, u}^\mathcal{G}},
\end{eqnarray}
\normalsize

\noindent
where $P^g_{Tx}$ is the transmission power of an omnidirectional antenna integrated in the \gNBs $g\!\in\!\mathcal{G}$;
$N_{g, u}$ represents thermal noise according to the allocated bandwidth; and most importantly, $I_{g,u}^\mathcal{G}$ is the interference level that user~$u$ suffers from other \gNBs.

\subsubsection{Ground-to-Air Channels}
\label{a:ss:G2A}

The aerial network relays traffic from the \gNBs by means of LoS backhaul wireless links. Hence, the attenuation of a \gNB--\aBS link $(g, a)$ is the following:

\small
\begin{eqnarray}
L_\mathcal{B}(g, a) = 10\eta_\mathcal{B} \log_{10} \left(\frac{4\pi f_\mathcal{B}}{c}\!\cdot\!  \text{dist}(g,a)\right) \!+\! \mathcal{N}\left(0, \sigma_\mathcal{B}^2\right),
  \label{eq:PL_G2A}
\end{eqnarray}
\normalsize

\noindent
where $\eta_\mathcal{B}\!\approx\! 2$ is the path-loss exponent in LoS; $f_\mathcal{B}$ is the operating carrier frequency of the backhaul wireless links; and $\sigma_\mathcal{B}^2$ is the standard deviation of the \blue{Gaussian} random variable $\mathcal{N}\left(0, \sigma_\mathcal{B}^2\right)$, modeling the effects of shadowing.

Backhaul links operate on the bandwidth shared with user access to \gNBs.
However, as backhaul links perform 3D-beamforming pointing to the air (where \aBSs hover), the interference between \gNB-served users and backhaul-served \aBSs is very limited. 
Although the majority of the \gNB radiating power is focused in one direction towards the air thanks to the adoption of 3D-beamforming,
non-ideal beam-patterns also radiate energy in other directions.
Therefore, the SINR experienced by an \aBS $a\!\in\!\mathcal{A}$ depends also on the direction in which other \gNBs transmit to other \aBSs. The SINR experienced by a \gNB--\aBS link $(g, a)$ is:

\small
\begin{eqnarray}
  \gamma^\mathcal{B}_{g, a} = \frac{P_{Tx}^g\!\cdot\! G_g\!\cdot\!  10^{-L_\mathcal{B}(g, a)/10}}{N^{g,a} + I^\mathcal{B}_{g, a}},
  \label{eq:SINRbackhaul}
\end{eqnarray}
\normalsize

\noindent
where $P_{Tx}^g$ is the transmission power of the \gNB $g$; $G_g$ is the antenna gain over the main lobe of the beam-pattern of \gNB $g$; $N^{g,a}$ is the thermal noise; and $I_{g, a}^\mathcal{B}$ is the interference coming from the remaining backhaul links of the network.

Backhaul links reuse the spectrum used for ground cellular connections, although using beam-patterns pointing to the air, while antennas that provide service to ground users are pointing mainly to the ground. Hence, we assume that the interference suffered by a backhaul link $(g, a)$ is dominated by the interference from other backhaul links.
Hence, the interference suffered by a backhaul link $(g, a)$ is:

\small
\begin{eqnarray}
I_{g, a}^\mathcal{B} = \sum\limits_{g'\in\mathcal{G}\setminus\{g\}} P_{Tx}^{g'} \!\cdot\!  G_{g'}(\phi_{g', a}) \!\cdot\!  10^{-L_\mathcal{B}(g', a)/10},
  \label{eq:Interf_Backhaul}
\end{eqnarray}
\normalsize

\noindent
where $\phi_{g', a}$ is the angle between the main lobe direction of the antenna of $g'$ and the position of \aBS $a$.
In case a \gNB $g'$ does no set any backhaul wireless link, this \gNB will not affect interference, and $P_{Tx}^{g'}$ will be considered as zero.

\end{document}